\documentclass[preprint,12pt,showpacs]{revtex4} 
\usepackage{amsmath}
\usepackage{latexsym}
\usepackage{amssymb}
\usepackage{graphics,epstopdf}
\usepackage{amsmath,amssymb,amsthm}
\usepackage{graphicx}
\usepackage[colorlinks=true, citecolor=blue, urlcolor=blue ]{hyperref}
\usepackage{slashed}
\theoremstyle{plain}

\begin{document}
\title{On the two-dimensional time-dependent anisotropic
harmonic oscillator in a magnetic field}

\author{Pinaki Patra}
\thanks{Corresponding author}
\email{monk.ju@gmail.com}
\affiliation{Department of Physics, Brahmananda Keshab Chandra College, Kolkata, India-700108}
\date{\today}

\begin{abstract}
 A Charged harmonic oscillator in a magnetic field, Landau problems, and an oscillator in a noncommutative space, share the same mathematical structure in their Hamiltonians. We have  considered a two-dimensional anisotropic harmonic oscillator (AHO) with arbitrarily time-dependent parameters  (effective mass and frequencies), placed in an arbitrarily time-dependent magnetic field. A class of quadratic invariant operators (in the sense of Lewis and Riesenfeld) have been constructed. The invariant operators ($\hat{\mathcal{I}}$) have been reduced to a simplified representative form by a linear canonical transformation (the
group $Sp(4, \mathbb{R})$). An orthonormal basis of the Hilbert space consisting of the eigenvectors of $\hat{\mathcal{I}}$ is obtained. In order to obtain  the solutions of the time-dependent Schr\"{o}dinger equation corresponding to the system, both the geometric and dynamical phase-factors are constructed. Generalized
Peres-Horodecki separability criterion (Simon’s criterion) for the ground state  corresponding to our system has been demonstrated.
\end{abstract}

 \maketitle
\section{Introduction}
 The long list of applications of a time-dependent harmonic oscillator (TDHO) in modelling a variety of physical phenomena makes the TDHO in a subject of active research \cite{tdho1,tdho2,tdho3,tdho4,tdho5,tdho6}. Dynamics of a  harmonic oscillator (HO) placed in a magnetic field is equivalent to the dynamics of a HO in a deformed noncommutative space \cite{ncs1,ncs2,ncs3,ncs4}, which is supposedly compatible with quantum theory of gravity \cite{qg1,qg2,qg3,qg4,qg5,qg6,qg7,qg8,qg9}. Moreover, Maxwell-Chern-Simons model in long wavelength limit and  a non-relativistic anyon in a magnetic field resembles with the two dimensional HO in a magnetic field \cite{mcs1,mcs2,mcs3,mcs4,mcs5,mcs6,mcs7,mcs8}. The systems with time-dependent parameters (mass, frequency, external magnetic fields etc.) appear to be self-righteous for the most physical systems, in particular for a system, which interacts with its environment \cite{tdho1,tdho2,tdho3,tdho4,tdho5,tdho6}. 
 For example, the  time-dependent effective mass (TDEM) appears for the energy density functional approach to many-body problems \cite{tdem1},  the study of electronic properties of condensed matter systems \cite{tdem2}, the Schr\"{o}dinger equation in curved space in the light of deformed algebras \cite{tdem3},  nonlinear optical properties in quantum well, the  cosmological models to quantum information theory \cite{tdem5}, the asymmetric shape of crackling noise pulses emitted by a diverse range of noisy systems can be modelled with the help of time-dependent effective mass \cite{tdem6}, and many more\cite{tdem7,tdem8}.
 \\
 The eigenstates of an annihilation operator (displaced ground state of HO), namely the coherent states (CS) are particularly important 
in interferometry
\cite{cs1,cs2,cs3,cs4}. In particular, the maximum contrast in the interference pattern is achieved for a CS \cite{cs5}. In the context of CS, a bipartite Gaussian states are particularly important for its frequent appearance in the systems with small oscillations \cite{cs6,cs7}. For example, vibration of the test mass (corresponding to the masses on which the interferometric mirrors are placed) under the effects of the tidal forces of gravitational waves are demonstrated by the anisotropic oscillators \cite{cs8,cs9,cs10}, which exhibits the bipartite Gaussian states upon consideration in a two-dimensional noncommutative space \cite{csncs1,csncs2,csncs3,csncs4}. \\ 
Therefore, an anisotropic harmonic oscillator is an important problem on its own right. In this article, we have considered a most general form of time-dependent anisotropic harmonic oscillator placed in a time-dependent external magnetic field. The time-dependent system is completely solved with the help of Lewis-Riesenfeld (LR) invariant method \cite{Lewis1,Lewis2,Lewis3,Lewis4,Lewis5,Lewis6,Lewis7}, which is based on the existence of an invariant operator $\hat{\mathcal{I}}(t)$, corresponding to the Hamiltonian $\hat{H}(t)$. If $\phi$ is an eigen-function of $\hat{\mathcal{I}}$ (i.e., $\hat{\mathcal{I}}\phi =\lambda \phi$), then $\psi=\phi e^{i\theta}$ will satisfy the time-dependent  Schr\"{o}dinger equation (TDSE) $i\hbar\partial  \psi/\partial t =\hat{H}\psi$, for some time-dependent phase factor $\theta(t)$ \cite{Lewis5,Lewis6,Lewis7}. LR-invariant (LRI) operators for time-dependent (TD) isotropic harmonic oscillators (HO) are well known in the literature \cite{ncs2}. On the other hand, time-independent anisotropic oscillators are farely well known \cite{ncs1,anisotropic1}. However, for arbitrary time-dependent anisotropic oscillator there is a gap in the literature. Our present study aims to fulfil this gap. We have constructed a quadratic invariant operator, which is transformed to a simpler form with the aid of a similarity transformation (symplectic group $Sp(4,\mathbb{R})$ \cite{anisotropic1,sp1}). The eigen-states and eigenvalues of $\hat{\mathcal{I}}$ are constructed by factorizing it with the annihilation and creation operators. The phase factors (both the geometrical and dynamical \cite{phase1,phase2,phase3,phase4}) are obtained in order to solve the TDSE. Then the separability conditions for the bipartite ground state (CS) have been illustrated with the generalized  Peres-Horodecki criterion (Simon's criterion) \cite{sep1,sep2,sep3,sep4,sep5,sep6}. \\
The organization of our paper is the followings. A brief discussion on our system is followed by the Lewis-Riesenfeld invariant method. An unitary transformation is constructed in order to transform the invariant operator in a simpler one, keeping the intrinsic symplectic structure intact. At last, the explicit form of the ground state is constructed and the explicit form of the separability condition for the concerned  state is mentioned.   
 
\section{Systems under consideration}
Hamiltonian of a charged particle inside a magnetic field can be expressed as a combination of the kinetic part (quadratic in momentum), along with a paramagnetic term (cross term of vector potential and momentum operator) and a diamagnetic contribution (quadratic in vector potential). Using the  Coulomb gauge condition ($\vec{\nabla}.\vec{A}=0$) of the magnetic vector potential $\vec{A}=(-\alpha_{01}(t)x_2,\alpha_{02}(t) x_1)$, 
 the paramagnetic term of the Hamiltonian for an anisotropic harmonic oscillator  with charge $e$ and effective mass  $M=(\mu_1(t),\mu_2(t))$ is reduced to
\begin{equation}
 \hat{L}_d = \nu_1 \left\{\hat{x}_2,\hat{p}_1\right\} - \nu_2 \left\{ \hat{x}_1,\hat{p}_2  \right\},
\end{equation}
whereas, the diamagnetic contribution is reduced to
\begin{equation}
 V_{diamag}=\frac{1}{2}(k_{02}x_1^2+ k_{01}x_2^2),
\end{equation}
where
\begin{equation}
 \nu_j=\frac{e}{2\mu_j }\alpha_{0j},\;\; k_{0j}=\frac{e^2 \alpha_{0j}^2}{\mu_j},\; j=1,2.
\end{equation}
Here we have set the speed of light in vacuum to unity, which will be followed throughout in this paper. The notation $\{ \mathcal{O}_1,\mathcal{O}_2\}=  \mathcal{O}_1\mathcal{O}_2 +  \mathcal{O}_2\mathcal{O}_1$ stands for the anticommutator of the operators $ \mathcal{O}_1$ and $\mathcal{O}_2$.\\
 Thus the Hamiltonian ($\hat{H}$) for an anisotropic harmonic oscillator (potential $V_{an}(x_1,x_2)=k_1(t)x_1^2/2 + k_2(t)x_2^2/2$) with charge $e$ and effective mass  $M=(\mu_1(t),\mu_2(t))$, under the magnetic vector potential ($-\alpha_{01}(t)x_2,\alpha_{02}(t) x_1$) reads
\begin{equation}\label{Hamiltonian}
 \hat{H}= \frac{1}{2\mu_1}\hat{p}_1^2 + \frac{1}{2\mu_2}\hat{p}_2^2 + \frac{1}{2}\alpha_1 \hat{x}_1^2 + \frac{1}{2}\alpha_2 \hat{x}_2^2 + \hat{L}_d ,
\end{equation}
Where 
\begin{equation}
    \alpha_1 = k_1+ 4\mu_2\nu_2^2, \; \alpha_2= k_2+ 4\mu_1\nu_1^2 .
\end{equation}
The system ~\eqref{Hamiltonian} is equivalent to an anisotropic oscillator (AHO) in a two dimensional noncommutative space (NC) ($[\hat{\tilde{X}}_1,\hat{\tilde{X}}_2]=i\theta ,
[\hat{\tilde{P}}_1,\hat{\tilde{P}}_2]=i\eta ,
 [\hat{\tilde{X}}_i,\hat{\tilde{P}}_j]=i\hbar_{e}\delta_{ij}; i,j=1,2.$), with the spatial NC parameter $\theta$ and momentum NC parameter $\eta$. Here $\hbar_e=\hbar(1+( \theta\eta)/(4\hbar^2))$, and $\delta_{ij}$ is Kronecker delta, which equals to one for $i=j$ and zero otherwise. The usual Heisenberg algebra for the commutative space ($
 [\hat{x}_i,\hat{x}_j]=[\hat{p}_i,\hat{p}_j]=0,\; [\hat{x}_i, \hat{p}_j] = i\hbar \delta_{ij}; i,j=1,2.
$) is connected with the NC-space algebra by the following Bopp's shift.
\begin{eqnarray}\label{boppshift}
 \hat{\tilde{X}}_j &=& \hat{x}_j+\frac{\theta}{2\hbar}\epsilon^{ij}\hat{p}_i,\;
\hat{ \tilde{P}}_j = \hat{p}_j + \frac{\eta}{2\hbar} \epsilon^{ji}\hat{x}_i;\;  i,j=1,2.
\end{eqnarray} 
Here $\epsilon^{ij}$ is completely antisymmetric tensor with $\epsilon^{12}=1$. With the transformations ~\eqref{boppshift}, it is evident  that the NC-space Hamiltonian 
\begin{equation}
 \hat{H}_{nc}=\frac{1}{2m_1}\hat{\tilde{P}}_1^2 + \frac{1}{2m_2}\hat{\tilde{P}}_2^2 + \frac{1}{2}m_1 \tilde{\omega}_1^2 \hat{\tilde{X}}_1^2 + \frac{1}{2}m_2 \tilde{\omega}_2^2\hat{\tilde{X}}_2^2 ,
\end{equation}
is equivalent to ~\eqref{Hamiltonian}, with the following identification of the time-dependent parameters.
\begin{eqnarray}
 \frac{1}{\mu_1} &=& \frac{1}{m_1}+ \frac{1}{4\hbar^2} m_2\tilde{\omega}_2^2\theta^2, \;
 \frac{1}{\mu_2} = \frac{1}{m_2}+ \frac{1}{4\hbar^2} m_1\tilde{\omega}_1^2\theta^2,\\
 \alpha_1&=& m_1\tilde{\omega}_1^2 + \frac{\eta^2}{4\hbar^2 m_2} ,\;
 \alpha_2  = m_2\tilde{\omega}_2^2 + \frac{\eta^2}{4\hbar^2 m_1},\\
 \nu_1 & =& \frac{1}{4m_1\hbar} (\eta + m_1 m_2\tilde{\omega}_2^2 \theta),\;
 \nu_2 = \frac{1}{4m_2\hbar} (\eta + m_1 m_2\tilde{\omega}_1^2 \theta).
\end{eqnarray}
The aim of the present paper is to study the Hamiltonian ~\eqref{Hamiltonian}, which  has the following quadratic form.
\begin{equation}\label{quadraticH}
 \hat{H}=\frac{1}{2}X^T \hat{\mathcal{H}}X,
\end{equation}
with 
\begin{eqnarray}
 X =  \left(\begin{array}{c}\hat{x}_1\\
  \hat{p}_1\\
   \hat{x}_2\\
    \hat{p}_2
    \end{array}
 \right),\; \mbox{and}\;\;
 \hat{\mathcal{H}} = \left(  \begin{array}{cccc} \label{Hquadratic}
\frac{\alpha_1}{2} & 0 & 0 & -\nu_2 \\
0 & \frac{1}{2\mu_1} & \nu_1 & 0 \\
0 & \nu_1 & \frac{\alpha_2}{2} & 0 \\
-\nu_2 & 0 & 0 &  \frac{1}{2\mu_2}
\end{array}
\right).
\end{eqnarray}
Here $T$ tenotes the matrix transposition.\\
One can see that the deformed angular momentum operator $\hat{L}_d$ will be reduced to the usual angular momentum operator for an isotropic oscillator (IHO). 
For an IHO,  the usual angular momentum operator ($\hat{L}$) commutes with the Hamiltonian.  Thus the problem can be reduced to the two noninteracting  oscillators. However, for AHO ($\nu_1\neq \nu_2$), the process of diagonalization is not starightforward. On the next section, we have constructed a class of invariant operators (in the sense of Lewis-Riesenfeld) corresponding to the time-dependent Hamiltonian ~\eqref{Hamiltonian}.
\section{Invariant operators}
At first we observe that, the set of operators 
\begin{eqnarray}
 \mathcal{A}= \left\{ \hat{x}_1^2, \hat{x}_2^2, \hat{p}_1^2, \hat{p}_2^2, \{\hat{x}_1,\hat{p}_1\}, \{\hat{x}_2,\hat{p}_2\}, \{\hat{x}_1,\hat{x}_2\}, \{\hat{p}_1,\hat{p}_2\}, \{\hat{x}_2,\hat{p}_1 \}, \{ \hat{x}_1, \hat{p}_2 \} \right\}
\end{eqnarray}
forms a closed quasi-algebra with respect to the Hamiltonian $\hat{H}$ as follows.
\begin{eqnarray}
 \left[ \hat{H}, \hat{x}_1^2\right] &=& -\frac{i\hbar}{\mu_1} \{\hat{x}_1,\hat{p}_1\} - i\hbar \nu_1 \{\hat{x}_1,\hat{x}_2 \}.
 \\
 \left[ \hat{H}, \hat{x}_2^2 \right] &=& -\frac{i\hbar}{\mu_2} \{\hat{x}_2,\hat{p}_2 \} + i\hbar \nu_2 \{\hat{x}_1,\hat{x}_2 \}.
 \\
\left[ \hat{H},\hat{p}_1^2 \right] &=& i\hbar\alpha_1 \{\hat{x}_1,\hat{p}_1\} -i\hbar \nu_2  \{\hat{p}_1,\hat{p}_2\}
  \\
\left[ \hat{H},\hat{p}_2^2 \right] &=& i\hbar\alpha_2 \{\hat{x}_2,\hat{p}_2 \} + i\hbar \nu_1  \{\hat{p}_1,\hat{p}_2\}
\end{eqnarray}
\begin{eqnarray}
\left[ \hat{H},\{\hat{x}_1,\hat{p}_1\} \right] &=& -\frac{2i\hbar}{\mu_1}\hat{p}_1^2 + 2i\hbar\alpha_1\hat{x}_1^2 -i\hbar (\nu_1\{\hat{x}_2,\hat{p}_1\} + \nu_2 \{\hat{x}_1,\hat{p}_2 \}) .
    \\
\left[ \hat{H},\{\hat{x}_2,\hat{p}_2 \} \right] &=& -\frac{2i\hbar}{\mu_2}\hat{p}_2^2 + 2i\hbar\alpha_2\hat{x}_2^2 + i\hbar (\nu_1\{\hat{x}_2,\hat{p}_1\} + \nu_2 \{\hat{x}_1,\hat{p}_2 \}) .
     \\
\left[ \hat{H},\{\hat{x}_1,\hat{x}_2\} \right] &=& -\frac{ i\hbar}{\mu_1} \{\hat{x}_2,\hat{p}_1\} - \frac{i\hbar}{ \mu_2}  \{\hat{x}_1,\hat{p}_2\} +2i\hbar (\nu_2\hat{x}_1^2 -\nu_1\hat{x}_2^2).
      \\
\left[ \hat{H},\{\hat{p}_1,\hat{p}_2\} \right] &=& i\hbar\alpha_1\{\hat{x}_1,\hat{p}_2 \} + i\hbar \alpha_2  \{\hat{x}_2,\hat{p}_1\} + 2i\hbar (\nu_1 \hat{p}_1^2 -\nu_2\hat{p}_2^2).
       \\
\left[ \hat{H},\{\hat{x}_2,\hat{p}_1\} \right] &=& -\frac{i\hbar}{\mu_2} \{\hat{p}_1,\hat{p}_2 \} + i\hbar \alpha_1 \{\hat{x}_1,\hat{x}_2 \} +i\hbar \nu_2 (\{\hat{x}_1,\hat{p}_1 \} -\{\hat{x}_2,\hat{p}_2 \}).
        \\
\left[ \hat{H},\{\hat{x}_1,\hat{p}_2 \} \right] &=& -\frac{i\hbar}{\mu_1} \{\hat{p}_1,\hat{p}_2 \} + i\hbar \alpha_2 \{\hat{x}_1,\hat{x}_2 \} +i\hbar \nu_1 (\{\hat{x}_1,\hat{p}_1 \} -\{\hat{x}_2,\hat{p}_2 \}).
\end{eqnarray}
The closed quasi-algebra of $\mathcal{A}$ suggests an ansatz for the invariant operator $\hat{\mathcal{I}}$ to be
\begin{eqnarray}\label{ansatzI}
 \hat{\mathcal{I}} = u_{11}\hat{x}_1^2 + u_{22}\hat{x}_2^2 + v_{11} \hat{p}_1^2 + v_{22} \hat{p}_2^2 + w_{11} \{\hat{x}_1,\hat{p}_1\}+ w_{22} \{\hat{x}_2,\hat{p}_2\} \nonumber \\ + u_{12} \{\hat{x}_1,\hat{x}_2\}+ v_{12} \{\hat{p}_1,\hat{p}_2\}+ w_{21} \{\hat{x}_2,\hat{p}_1 \} + w_{12} \{ \hat{x}_1, \hat{p}_2 \}. 
\end{eqnarray}
The invariance of $\hat{\mathcal{I}}$ implies
\begin{equation}\label{invarianceI}
 \frac{d\hat{\mathcal{I}}}{dt}= \frac{\partial \hat{\mathcal{I}}}{\partial t} + \frac{1}{i\hbar} \left[ \hat{\mathcal{I}}, \hat{H}\right] =0.
\end{equation}
Using the ansatz ~\eqref{ansatzI} in ~\eqref{invarianceI} we get the following set of coupled equations of the coefficients.
\begin{eqnarray}
 \dot{w}= \nu v . \label{wdot}\\
 \dot{u}= 2\alpha v . \label{udot}\\
 \dot{v}= \beta u + \mu w. \label{vdot}
\end{eqnarray}
The vectors $w,u,v$ are given by
\begin{eqnarray}
 w = (w_{12}, w_{21})^T,\;
 u = (u_{11}, v_{11}, u_{22}, v_{22})^T , \;
 v = (v_{12}, w_{22}, w_{11}, u_{12})^T .
 \end{eqnarray}
 The coefficient matrices are as follows.
 \begin{eqnarray}
\mu = \left( \begin{array}{cc}
-\frac{1}{\mu_1} & -\frac{1}{\mu_2} \\
-\nu_1 & -\nu_2 \\
\nu_1 & \nu_2 \\
\alpha_2 & \alpha_1
\end{array}
\right),\;\;
\alpha = \left( \begin{array}{cccc}
0 & 0 & \alpha_1 & \nu_2 \\
\nu_1 & 0 & -\frac{1}{\mu_1} & 0 \\
0 & \alpha_2 & 0 & -\nu_1 \\
-\nu_2 & -\frac{1}{\mu_2} & 0 & 0
\end{array}
\right).\\
\nu = \sigma_x \mu^T \mathcal{S}_x ,\;\;
\beta = \left(\begin{array}{cccc}
0 & -\nu_2 & 0 & \nu_1 \\
0 & 0 & -\frac{1}{\mu_2} & \alpha_2 \\
-\frac{1}{\mu_1} & \alpha_1 & 0 & 0 \\
-\nu_1 & 0 & \nu_2 & 0
\end{array}\right).
\end{eqnarray}
With
\begin{eqnarray}
 \mathcal{S}_x =\left(\begin{array}{cc}
0 & \sigma_x \\
\sigma_x & 0
\end{array}
\right),\; \mbox{and}\; \sigma_x =\left(\begin{array}{cc}
0 & 1 \\
1 & 0
\end{array}
\right).
\end{eqnarray}
For a given set of time dependent  parameters, one can solve the above set of equations ~\eqref{wdot}, ~\eqref{udot} and ~\eqref{vdot}, in principle. \\
The invariant operator ~\eqref{ansatzI} is transformed into a simple form on the next section.

\section{Diagonalization of the system}
We can express the invariant operator ~\eqref{ansatzI} in the following quadratic form.

\begin{equation}
 \hat{\mathcal{I}}=\frac{1}{2}X^T \hat{\mathcal{H}}_{\mathcal{I}}X,
\end{equation}
with
\begin{eqnarray}
 \hat{\mathcal{H}}_\mathcal{I} = \left( \begin{array}{cc}
 \hat{C} & \hat{A}^T \\
 \hat{A} & \hat{B}
\end{array}\right).
\end{eqnarray}
Where
\begin{eqnarray}
 \hat{C} = \left( \begin{array}{cc}
 u_{11} & w_{11} \\
 w_{11} & v_{11}
\end{array}\right),
\hat{B} = \left( \begin{array}{cc}
u_{22} & w_{22} \\
 w_{22} & v_{22}
\end{array}\right),
\hat{A}= \left( \begin{array}{cc}
u_{12} & w_{21} \\
 w_{12} & v_{12}
\end{array}\right).
\end{eqnarray}
One can identify that
\begin{eqnarray}\label{symplecticdefn}
 \hat{\mathcal{S}}\hat{\mathcal{J}}_4 + \hat{\mathcal{J}}_4 \hat{\mathcal{S}}^T =0,
\end{eqnarray}
where
\begin{eqnarray}\label{symplecticmatrix}
 \hat{\mathcal{S}}= \hat{\mathcal{J}}_4\hat{\mathcal{H}}_\mathcal{I}, \; \mbox{with} \;\;
 \hat{\mathcal{J}}_{2n}= \left( \begin{array}{cc}
0 & \hat{\mathbb{I}}_n \\
-\hat{\mathbb{I}}_n & 0
\end{array}\right);
\end{eqnarray}
$\hat{\mathbb{I}}_n$ being the $n\times n$ identity matrix.
From the equation~\eqref{symplecticdefn}, we can conclude that
$\mathcal{S} \in sp (4,\mathbb{R})$. 
Since our co-ordinates are the Cartesian type (the natural range
(spectrum) for each of them is the entire real line),  the commutation relations 
\begin{equation}\label{columncommutation}
 [\hat{X}_\alpha, \hat{X}_\beta] =-( \Sigma_y)_{\alpha\beta};\; (\alpha,\beta=1,2,3,4),
\end{equation}
generates an intrinsic symplectic matrix
\begin{eqnarray}\label{Sigmay}
 \hat{\Sigma}_y=\mbox{diag}(\sigma_y,\sigma_y).
\end{eqnarray}
Here we have used the unit such that $\hbar=1$, which will be followed throughout the present paper, unless otherwise specified.
$\hat{\sigma}_y$  being the second Pauli matrix
\begin{eqnarray}
 \sigma_y =\left(\begin{array}{cc}
0 & -i \\
i & 0
\end{array}
\right).
\end{eqnarray}
The aim of the present section is to diagonalize $\hat{\mathcal{H}}_\mathcal{I}$ keeping the symplectic structure ~\eqref{columncommutation} intact.\\
The structure constants of the closed quasi-algebra of the co-ordinates ($\hat{X}_i,i=1,..,4.$) with respect to the Hamiltonian ($\hat{H}$), induce the following commutation relation.
\begin{equation}
 \frac{1}{2\hbar}\left[i \hat{\mathcal{I}}, X\right] =\hat{\Omega} X.
\end{equation}
Where
\begin{equation}\label{OmegaHconnection}
 \hat{\Omega}= \left(\begin{array}{cccc}
w_{11} & v_{11} & w_{21 } & v_{12} \\
-u_{11} & - w_{11} & -u_{12} & -w_{12}\\
w_{12} & v_{12} & w_{22} & v_{22}\\
-u_{12} & -w_{21} & -u_{22} & -w_{22}
\end{array}
\right).
\end{equation}
We observe the followings.
\begin{eqnarray}
 \hat{\Omega}= \left(\begin{array}{cc}
A_1 & B_1 \\
C_1 & D_1
\end{array}
\right),
\end{eqnarray}
where
\begin{eqnarray}
 A_1= \left(\begin{array}{cc}
w_{11} & v_{11} \\
-u_{11} & -w_{11}
\end{array}
\right), \;
B_1= \left(\begin{array}{cc}
w_{21} & v_{12} \\
-u_{12} & -w_{12}
\end{array}
\right),\\
C_1= \left(\begin{array}{cc}
w_{12} & v_{12} \\
-u_{12} & -w_{21}
\end{array}
\right), \; 
D_1= \left(\begin{array}{cc}
w_{22} & v_{22} \\
-u_{22} & -w_{22}
\end{array}
\right).
\end{eqnarray}
Since
\begin{eqnarray}
 A_1 = i\sigma_y C ,\;
 C_1 = i\sigma_y A,\;
 D_1= i\sigma_y B,
\end{eqnarray}
we have 
\begin{eqnarray}
 \Delta_{A_1} = \Delta_{C},\;
 \Delta_{B_1} = \Delta_{A},\\
 \Delta_{C_1} =\Delta_{A},\;
 \Delta_{D_1} = \Delta_{B}.
\end{eqnarray}
Where the notation $\Delta_{Q}$ denotes the determinant of the matrix $Q$.\\
We  wish to diagonalize $\hat{\Omega}$, and construct an equivalent co-ordinate system for our purpose to be served. The  construction of  $\hat{Q}^{-1}$ and $\hat{Q}$ can be done by arranging the normalized eigenvectors ( left and right) column-wise.
Although, $\hat{\Omega}$ is a normal matrix $
 ([ \hat{\Omega}^\dagger ,\hat{\Omega}] =0
$, it is not a symmetric one
$
 (\hat{\Omega}^T \neq \hat{\Omega})
$). Therefore, the left and right eigen-vectors of $\hat{\Omega}$ are not same (although, the left and right eigen-values are the same). 
The characteristic polynomial 
($
 p(\lambda)= \det (\lambda \mathbb{\hat{I}}- \hat{\Omega}) ,
$)
of  $\hat{\Omega}$ is given by
\begin{equation}
 p(\lambda)= \lambda^4+ \Delta \lambda^2 +\Delta_{\Omega} ,\; \mbox{with}\;\;
\Delta = \Delta_C + \Delta_B + 2\Delta_A.
\end{equation}
One can see that $p(\lambda)$ has four purely imaginary roots for the possible time-dependent real parameters in our problem.
\begin{equation}
 \lambda \in \{\pm i\sigma_j,\; j=1,2\vert \sigma_j^2=\frac{1}{2}(\Delta \pm \sqrt{\Delta^2 -4\Delta_\Omega}) \}.
\end{equation}
Let the eigenvalue equations for $\hat{\Omega}$ are given as follows.
\begin{eqnarray}
 \chi_{lj} \hat{\Omega} &=& -i\sigma_j \chi_{lj},\; j=1,2.\\
 \chi_{lj}^* \hat{\Omega} &=& i\sigma_j \chi_{lj}^*\;, j=1,2.\\
 \hat{\Omega} \chi_{rj} &=& -i\sigma_j \chi_{rj},\; j=1,2. \\
 \hat{\Omega}\chi_{rj}^* &=& i\sigma_j \chi_{rj}^*,\; j=1,2.
\end{eqnarray}
Where the suffix $l$ and $r$ stands for the left and right eigen-vectors respectively.
The following identities hold quite generally
\begin{eqnarray}
 \chi_{lj}^* \chi_{rj} &=& \chi_{lj} \chi_{rj}^*,\forall i,j=1,2.\\
 \chi_{lj} \chi_{rj} &=& \chi_{lj}^* \chi_{rj}^* =\delta_{ij}.
\end{eqnarray}
This suggests the relationship between the left and right eigenvalues as
\begin{equation}\label{leftrightconnection}
 \chi_{rj}= -\Sigma_y \chi_{lj}^\dagger, \; (j=1,2),
\end{equation}
which can be used as the normalization conditions.
In our case, the explicit form of the left eigenvectors $\chi_{lj},\;(j=1,2)$ are given by
\begin{eqnarray}
 \chi_{lj} = \frac{1}{k_j} \left(
 s_{j1}+iq_{j1}, s_{j2}+iq_{j2}, s_{j3}, s_{j4}+iq_{j4}
\right).
\end{eqnarray}
Where
\begin{eqnarray}
 s_{j1} &=& u_{11}^2 (v_{12}u_{22}-w_{21}w_{22}) + u_{11}u_{12}(\sigma_j^2 +w_{12}w_{21} \nonumber \\ 
 && +w_{11}w_{22}-v_{12}u_{12})  - u_{11}u_{22}w_{11}w_{12}. \label{sj1}\\
 q_{j1}&=& \sigma_j u_{11}(w_{12}u_{22}- u_{12}w_{22}-u_{11}w_{21}+u_{12}w_{11}).\\
 s_{j2} &=& \sigma_j^2 (u_{11}w_{21} + u_{12}w_{22} - u_{22}w_{12}) +w_{11}^2 (u_{12}w_{22} - u_{22}w_{12}) \nonumber \\ 
 && + w_{21} w_{11} ( u_{12}w_{12} - u_{11}w_{22}) + w_{11} v_{12} (u_{11}u_{22} - u_{12}^2). \\
 q_{j2} &=& \sigma_j^3 u_{12} + \sigma_j [u_{12}(w_{12}w_{21}-v_{12}u_{12}+w_{11}^2) + u_{11} (v_{12}u_{22} - w_{21}w_{22} -w_{11} w_{21})].\\
 s_{j3} &=& u_{11} u_{22} (\sigma_j^2 + w_{11}^2 - u_{11} v_{11}) + u_{11} u_{12} (v_{11} u_{12}  - w_{11} w_{21}) \nonumber \\ 
 && + u_{11}w_{21} ( u_{11} w_{21} - u_{12} w_{11}).\\
 s_{j4} &=& u_{11}w_{22} \sigma^2 + u_{11}^2 (v_{12}w_{21} - v_{11}w_{22}) + u_{12}u_{11} (v_{11}w_{12} - v_{12}w_{11}) \nonumber \\ 
 && + u_{11}w_{11} (w_{11}w_{22} - w_{12}w_{21}). \\
 q_{j4} &=& u_{11}\sigma_1^3 + u_{11}\sigma_1 (w_{11}^2 - u_{11}v_{11} + w_{21}w_{12} -v_{12}u_{12}). \label{qj4}
\end{eqnarray}
$k_j$ is the normalization factor. \\
For the time-independent Hamiltonian for an anisotropic harmonic oscillator we have
\begin{equation}
 w_{11}=w_{22}=u_{12}=v_{12}=0.
\end{equation}
In that case the components of the left eigenvectors simplifies to
\begin{eqnarray}
 s_{j1}=q_{j2}=s_{j4}=0.\\
 q_{j1}=\sigma_j u_{11}(w_{12}u_{22}- u_{11}w_{21}).\\
 s_{j2} = \sigma_j^2 (u_{11}w_{21}- u_{22}w_{12}).\\
 s_{j3}= u_{11}u_{22} \sigma_j^2 + u_{11}^2 (w_{21}^2 -v_{11}u_{22}).\\
 q_{j4}= u_{11}\sigma_1^3 + u_{11}\sigma_1 (w_{21}w_{12}- u_{11}v_{11}).
\end{eqnarray}
That means, for time-independent case, the components are either real or purely imaginary, which simplifies the calculations.\\
However, for general time-dependent systems, we have to consider the components ~\eqref{sj1} to ~\eqref{qj4}.
The right eigenvectors $(\chi_{rj},\; j=1,2)$ can be constructed from ~\eqref{leftrightconnection} in a straighforward manner. \\
Now the similarity transformation $\hat{Q}$ (as well as $\hat{Q}^{-1}$) can be constructed by arranging the  eigen-vectors columnwise. In particular, 
\begin{eqnarray}
 \hat{Q} &=& (\chi_{r1}, \chi_{r1}^*, \chi_{r2}, \chi_{r2}^*), \\
 \hat{Q}^{-1} &=& (\chi_{l1}^T, (\chi_{l1}^*)^T, \chi_{l2}^T, (\chi_{l2}^*)^T).
\end{eqnarray}
One can  verify that 
\begin{equation}\label{Qdagger}
 \hat{Q}^\dagger = -\hat{\Sigma}_z \hat{Q}^{-1} \hat{\Sigma}_y,
\end{equation}
with
\begin{equation}
 \hat{\Sigma}_z = \mbox{diag}(\sigma_z,\sigma_z).
\end{equation}
Where $\sigma_z$ is the third Pauli matrix
\begin{eqnarray} 
\sigma_z= \left(\begin{array}{cc}
1 &0\\
0&-1
\end{array}
\right).
\end{eqnarray}
Now, the diagonalization gives
\begin{equation}
 \hat{Q}_D= \hat{Q}^{-1} \hat{\Omega}\hat{Q}=\mbox{diag}(-i\sigma_1,i\sigma_1,-i\sigma_2,i\sigma_2).
\end{equation}
Whether this diagonalization helps our purpose, or not, can be seen by the followings. \\
Let the new co-ordinate vector $\hat{a}$ is given by 
\begin{equation}\label{zetadefn}
 \hat{a}= \hat{Q}^{-1}\hat{\zeta}.
\end{equation}
Then, using ~\eqref{Qdagger}, we can write
\begin{eqnarray}\label{zetadagger}
 \therefore \zeta^\dagger &=&  \hat{a}^\dagger (-\hat{\Sigma}_z \hat{Q}^{-1} \hat{\Sigma}_y) .
\end{eqnarray}
Then, the time-independent eigen-value equation for the invariant operator $\hat{\mathcal{I}}$
\begin{equation}
 \hat{\mathcal{I}}\psi =\lambda \hat{\mathcal{I}},
\end{equation}
is transformed as
\begin{eqnarray}
 \hat{Q}^{-1} \frac{1}{2} \zeta^T \hat{\mathcal{I}} \zeta \psi \hat{Q} &=& \lambda \hat{Q}^{-1} \hat{\mathcal{I}} \hat{Q} \\
 \therefore \frac{1}{2} \hat{a}^\dagger \Sigma \hat{a} \psi &=& \lambda \hat{\Sigma} \psi.
\end{eqnarray}
Where
\begin{equation}\label{Hdiag}
 \hat{\mathcal{I}}_D=\hat{\Sigma}
  = \mbox{diag}(\sigma_1,\sigma_1,\sigma_2,\sigma_2).
\end{equation}
That means, the new equivalent diagonal Hamiltonian $\hat{\Sigma}$ corresponding to the invariant operator $\hat{\mathcal{I}}$
obeys the same dynamics as that of our original one ($\hat{\mathcal{I}}$).

\section{Eigen-functions and eigen-values of the invariant operator}
If we compute the commutators of the components of $\hat{a}$, then we can identify that they forms a set of independent annihilation and creation operators. In particular, we can identify that
\begin{eqnarray}
 \hat{\zeta}^T =(\hat{a}_1, \hat{a}_1^\dagger, \hat{a}_2, \hat{a}_2^\dagger),
\end{eqnarray}
with the following commutation relations.
\begin{eqnarray}
 [\hat{a}_k, \hat{a}_l^\dagger]=\delta_{kl}, \; [\hat{a}_k, \hat{a}_l]= [\hat{a}_k^\dagger, \hat{a}_l^\dagger]=0;\; k,l=1,2.
\end{eqnarray}
The eigenstates of $\hat{\mathcal{I}}$ can be expressed as
\begin{eqnarray}
 \vert n_1,n_2\rangle = \frac{1}{\sqrt{n_1!n_2!}} (\hat{a}_1^\dagger)^{n_1} (\hat{a}_2^\dagger)^{n_2}\vert 0,0\rangle ; n_1,n_2=0,1,2,3,.....
\end{eqnarray}
The corresponding energy eigenvalues are
\begin{equation}
 E_{n_1,n_2}=\left(n_1+\frac{1}{2}\right)\sigma_1 + \left(n_2+\frac{1}{2}\right)\sigma_2.
\end{equation}
\section{Time-dependent Phase factor and the solutions of Schr\"{o}dinger equation}
If $\vert n_1,n_2\rangle$ is an eigenstate of the invariant operator $\mathcal{\hat{I}}$,  then the time-dependent Schr\"{o}dinger equation will be satisfied by the same function multiplied with some time-dependent phase factor ($\theta_{n_1,n_2}(t)$). In particular,
\begin{equation}
 \hat{H} \left( \vert n_1,n_2\rangle e^{i\theta_{n_1,n_2}(t)} \right) = i \frac{\partial}{\partial t} \left( \vert n_1,n_2\rangle e^{i\theta_{n_1,n_2}(t)} \right) .
\end{equation}
That means
\begin{equation}
  \dot{\theta}_{n_1, n_2} = \langle n_1,n_2 \vert \left( i \frac{\partial}{\partial t} - \hat{H} \right) \vert n_1, n_2 \rangle .
\end{equation}
We can split the phase-factor into geometric phase $\theta^g $ (including the Berry 
phase phenomenon for slowly varying parameters) and dynamical phase $\theta^d$ as follows.
\begin{eqnarray}
 \dot{\theta}_{n_1,n_2}^g = i \langle n_1,n_2 \vert \frac{\partial}{\partial t} \vert n_1, n_2 \rangle. \label{geometrical} \\
  \dot{\theta}_{n_1,n_2}^d = - \langle n_1,n_2 \vert \hat{H} \vert n_1, n_2 \rangle. \label{dynamical}
\end{eqnarray}
For the ground state of time-dependent anisotropic oscillator
\begin{equation}
 \dot{\theta}_{0,0}^g =\frac{\dot{N}_0}{N_0} - \dot{a}\langle \hat{x}_1^2 \rangle - \dot{b}\langle \hat{x}_2^2 \rangle - \dot{c}\langle \hat{x}_1\hat{x}_2 \rangle .
\end{equation}
Using the equations ~\eqref{zetadefn} and ~\eqref{zetadagger}, the quadratic form ~\eqref{quadraticH} can be rewritten as
\begin{equation}\label{diagonalah}
 \hat{H}= \hat{a}^\dagger \left(   - \Sigma_z \hat{Q}^{-1} \Sigma_y \right) \hat{\mathcal{H}} \hat{Q} \hat{a}.
\end{equation}
Since
\begin{equation}
 \tilde{\Omega}= \Sigma_y \mathcal{H},
\end{equation}
is a subclass of the matrix $\Omega$, it will be diagonalized by $Q$. The characteristic polynomial ($P_{\tilde{\Omega}}(\tilde{\lambda})$) of $\tilde{\Omega}$ is
\begin{equation}
P_{\tilde{\Omega}}(\tilde{\lambda}) = \tilde{\lambda}^4 + \tilde{\Delta}\tilde{\lambda}^2 + \Delta_{\tilde{\Omega}}, 
\end{equation}
with
\begin{eqnarray}
 \tilde{\Delta} &=& -\frac{\alpha_1}{4\mu_1} -\frac{\alpha_2}{4\mu_2} - 2\nu_1\nu_2,\\
  \Delta_{\tilde{\Omega}} &=& \frac{\alpha_1 \alpha_2}{16 \mu_1\mu_2} -\frac{\nu_2^2 \alpha_2}{4\mu_1} - \frac{\nu_1^2 \alpha_1}{4\mu_2} +\nu_1^2 \nu_2^2.
\end{eqnarray}

$P_{\tilde{\Omega}}(\tilde{\lambda})$ has positive discriminant
\begin{equation}
 D_{\tilde{\Omega}}= \left( \frac{\alpha_1}{4\mu_1} - \frac{\alpha_2}{4\mu_2}\right)^2 + (\alpha_1\nu_1 +\alpha_2\nu_2) \left( \frac{\nu_1}{\mu_2} + \frac{\nu_2}{\mu_1}\right) \ge 0.
\end{equation}
In other words, $\tilde{\lambda}^2$ is real. Moreover,
\begin{equation}
 \tilde{\lambda}^2 =\frac{1}{2}\left(-\tilde{\Delta}+ \sqrt{D_{\tilde{\Omega}}}\right) \ge 0,
\end{equation}
gives two real roots for $\tilde{\lambda}$, which reads
\begin{equation}
 \tilde{\sigma}_{\pm}= \pm \frac{1}{\sqrt{2}} \sqrt{-\tilde{\Delta}+ \sqrt{D_{\tilde{\Omega}}}} =\pm \tilde{\sigma}_1.
\end{equation}
Other two roots will be either real or purely imaginary depending on the parameter values. In particular, 
\begin{eqnarray}\label{parameters}
 (\alpha_2 -4\mu_1\nu_1^2)(\alpha_1 -4\mu_2\nu_2^2)\ge 0 \implies \tilde{\lambda}^2 \ge  0, \\
 (\alpha_2 -4\mu_1\nu_1^2)(\alpha_1 -4\mu_2\nu_2^2)\le 0 \implies \tilde{\lambda}^2 \le 0.
\end{eqnarray}
It is worth noting that, for an  isotropic oscillator $\tilde{\lambda}^2$ will  be positive for all parameter values. For our present case (anisotropic),  the roots 
\begin{equation}
\pm  \tilde{\sigma}_2 = \pm \frac{1}{\sqrt{2}} \sqrt{-\tilde{\Delta}-  \sqrt{D_{\tilde{\Omega}}}},
\end{equation}
are either real or purely imaginary, depending on the parameter values ~\eqref{parameters}.
The ~\eqref{diagonalah} is reduced to the following.
\begin{equation}\label{hdiagonaltilde}
 \hat{H}=-\tilde{\sigma}_1 ( 2 \hat{a}_1^\dagger \hat{a}_1 +1 ) - \tilde{\sigma}_2 (2 \hat{a}_2^\dagger \hat{a}_2 + 1 )
\end{equation}
Then the dynamical phase ~\eqref{dynamical} can be expressed as
\begin{equation}
  \dot{\theta}_{n_1,n_2}^d = (2n_1 +1)\tilde{\sigma}_1 + (2n_2 +1)\tilde{\sigma}_2 .
\end{equation}
For the ground state
\begin{equation}
  \dot{\theta}_{0,0}^d = \sqrt{-\tilde{\Delta}+ 2\sqrt{\Delta_{\tilde{\Omega}}}}.
\end{equation}

Given the time dependent parameters, the first order ordinary differential equations for the dynamical and geometrical phases can be integrated in a straightforward manner.

\section{Separability of bipartite ground states}
Given a bipartite density operator $\hat{\rho}$, we can define
\begin{equation}
 \Delta \hat{X}= \hat{X}- \langle \hat{X}\rangle,\;\;\mbox{where}\; \langle \hat{X}_\alpha \rangle =Tr(\hat{X}_\alpha \hat{\rho}),\; \alpha=1,2,3,4.
\end{equation}
The components of $\Delta \hat{X}$ satisfy the commutation relation ~\eqref{columncommutation} as of $\hat{X}$. Uncertainties (variances) may be defined as expectation velues of the Hermitian operators
\begin{equation}
 \hat{\mathcal{V}}_{\alpha\beta} =\frac{1}{2}\{\Delta \hat{X}_\alpha, \Delta \hat{X}_\beta \},\; \alpha,\beta=1,2,3,4,
\end{equation}
which enables us to define the variance matrix $\hat{V}$ through the matrix elements
\begin{equation}\label{valphabeta}
 \mathcal{V}_{\alpha\beta}=\langle \hat{\mathcal{V}}_{\alpha\beta} \rangle = \frac{1}{2}\langle \{ \hat{X}_\alpha, \hat{X}_\beta \} \rangle -\langle \hat{X}_\alpha \rangle \langle \hat{X}_\beta \rangle ,\; \alpha,\beta=1,2,3,4.
\end{equation}
The statement of the uncertainty principle is then given by
\begin{equation}\label{uncert1}
 \hat{V}- \frac{1}{2}\hat{\Sigma}_y \ge 0,
\end{equation}
which has $Sp(4,\mathbb{R})$ invariant form.  Under the
Peres-Horodecki partial transpose ($\hat{\mathcal{T}}$), the Wigner distribution $ (W(x_1,p1,x_2,p_2)= \pi^{-2}\int dx_1'dx_2' \langle x-x'\vert \hat{\rho}\vert x+x'\rangle \exp(2i\vec{x}'.\vec{p})) $
undergoes mirror reflection ($\hat{\mathcal{T}}: W(x_1,p1,x_2,p_2)\to W(x_1,p1,x_2,-p_2)$, with $\hat{\mathcal{T}}=diag(1,1,1,-1)$). Peres-Horodecki criterion states that if the bipartite state is separable then $W(\hat{\mathcal{T}}X)$ has to be Wigner distribution. In other words, for a separable state, after the partial transpose operation, the transformed variance matrix $\hat{\tilde{V}}$ should satisfy same uncertainty relation i.e., $\hat{\tilde{V}} - \frac{1}{2}\hat{\Sigma}_y$ should be a nonnegative matrix
\begin{equation}\label{uncert2}
 \hat{\tilde{V}}- \frac{1}{2}\hat{\Sigma}_y \ge 0,\;\mbox{with}\; \hat{\tilde{V}}= \hat{\mathcal{T}} \hat{V}\hat{\mathcal{T}}.
\end{equation}
~\eqref{uncert1} and ~\eqref{uncert2} is preserved only under $Sp(2,\mathbb{R})\bigotimes Sp(2,\mathbb{R})$ subgroup of $Sp(4,\mathbb{R})$ corresponding to independent local linear canonical transformations $\hat{S}_1$ and $\hat{S}_2$ on the subsystems of two independent observers respectively. In other words, $\hat{S}_{local}=diag(\hat{S}_1,\hat{S}_2)$ with $\hat{S}_1\mathcal{J}_2\hat{S}_1^T =\hat{S}_2\mathcal{J}_2\hat{S}_2^T= \mathcal{J}_2$, where $ \mathcal{J}_2$ is given in ~\eqref{symplecticmatrix}.
On the other hand, since $[\hat{x}_j,\hat{p}_k]=i\delta_{jk}$, from ~\eqref{valphabeta}, we can see that $\hat{V}$ can be written as the following block form
\begin{eqnarray}
 \hat{V}= \left(\begin{array}{cc}
 \hat{V}_{11} & \hat{V}_{12} \\
\hat{V}_{12}^T & \hat{V}_{22}
\end{array}
\right).
\end{eqnarray}
The local groups transform $2\times 2$ blocks as follows.
\begin{equation}
 \hat{V}_{11}\to \hat{S}_1 \hat{V}_{11}\hat{S}_1^T,\; \hat{V}_{22}\to \hat{S}_2 \hat{V}_{22}\hat{S}_2^T,\; \hat{V}_{12}\to \hat{S}_1 \hat{V}_{12}\hat{S}_2^T.
\end{equation}
Therefore, the $Sp(2,\mathbb{R})\bigotimes Sp(2,\mathbb{R})$ invariants corresponding to $\hat{V}$ are  
\begin{equation}
 \Delta_1=Det(\hat{V}_{11}),\;  \Delta_2=Det(\hat{V}_{22}),\;  \Delta_{12}=Det(\hat{V}_{12}),\; \tau_V=\mbox{Trace}(\hat{V}_{11}\hat{\tilde{V}}_{12}\hat{V}_{22}\hat{\tilde{V}}_{21}) .
\end{equation}
Where 
\begin{eqnarray}
 \hat{\tilde{V}}_{jk}= \mathcal{J}_2 V_{jk}\mathcal{J}_2, \; \mbox{with}\; 
 \hat{V}_{21}=\hat{V}_{12}^T.
\end{eqnarray}
It was shown by Simon \cite{sep1} that the necessary and sufficient condition for separability condition ( generalized Peres-Horodecki criterion) is equivalent to the following inequality 
\begin{eqnarray}\label{simonscriterion}
 \Delta_1\Delta_2 + \left(\frac{1}{4}- \vert \Delta_{12}\vert\right)^2 -\mbox{Trace}(V_{11}\tilde{V}_{12}V_{22}\tilde{V}_{21}) \ge (\Delta_1+ \Delta_2).
\end{eqnarray}
To construct the exact form of the inequality corresponding to our system  we need to construct the state of our system under consideration. We shall restrict ourselves only to the ground state, which is given by
\begin{equation}\label{groundstateequation}
 \hat{a}_1\vert 0,0\rangle =\hat{a}_2\vert 0,0\rangle =0.
\end{equation}
In the position ($\left\{\vert x\rangle\right\}$) representation, equation ~\eqref{groundstateequation} reads
\begin{eqnarray}\label{xrepresentationequation}
 \left(\Lambda_1^{(l)}x - i\hbar \Lambda_2^{(l)}\partial_x\right)\psi_{00}=0.
\end{eqnarray}
Where
\begin{eqnarray}
 \Lambda_1^{(l)} = \left( \begin{array}{cc}
\chi_{l11} & \chi_{l13}\\
\chi_{l21} & \chi_{l23}
\end{array}
\right) ,\;\;
\Lambda_2^{(l)} = \left( \begin{array}{cc}
\chi_{l12} & \chi_{l14}\\
\chi_{l22} & \chi_{l24}
\end{array}
\right),
\end{eqnarray}
with
\begin{equation}
 \chi_{lj}=(\chi_{lj1}, \chi_{lj2}, \chi_{lj3}, \chi_{lj4}),\; j=1,2.
\end{equation}
The notations $
 x=(x_1,x_2)^T,\;
\partial_x=(\frac{\partial}{\partial x_1},\frac{\partial}{\partial x_2})^T$ have been used.

Let us consider the following ansatz for the solution of ~\eqref{xrepresentationequation}.
\begin{equation}\label{ansatzpsi00}
 \psi_{00}= N_0 e^{-S(x_1,x_2)},
\end{equation}
with
\begin{equation}
 S(x_1,x_2)= \frac{1}{2}x^T \Lambda x .
\end{equation}
Using ~\eqref{ansatzpsi00} in ~\eqref{xrepresentationequation}, we get
\begin{equation}
 \Lambda = i(\Lambda_2^{(l)})^{-1} \Lambda_1^{(l)}.
\end{equation}
The explicit form in the $\{\vert x\rangle\}$ representation ($\psi(x_1,x_2)=\langle x_1,x_2\vert 0,0\rangle$) reads
\begin{equation}\label{psientangle}
 \psi(x_1,x_2)=N_0 e^{-\frac{1}{2}(\Lambda_{11}x_1^2+\Lambda_{22}x_2^2+ (\Lambda_{12} +\Lambda_{21})x_1x_2)},
\end{equation}
$N_0$ being the normalization constant. The matrix elements of $\Lambda$ are as follows.
\begin{eqnarray}
 \Lambda_{11} &=& \frac{i}{ \Delta_{\Lambda_2^{(l)}}} \left( \chi_{l24}\chi_{l11} - \chi_{l14} \chi_{l21} \right).\\
 \Lambda_{12} &=& \frac{i}{ \Delta_{\Lambda_2^{(l)}}} \left( \chi_{l24}\chi_{l13} - \chi_{l14} \chi_{l23} \right). \\
 \Lambda_{21} &=& \frac{i}{ \Delta_{\Lambda_2^{(l)}}} \left( \chi_{l12}\chi_{l21} - \chi_{l22} \chi_{l11} \right). \\
 \Lambda_{22} &=& \frac{i}{\Delta_{\Lambda_2^{(l)}}} \left( \chi_{l12}\chi_{l23} - \chi_{l22} \chi_{l13} \right).
\end{eqnarray}
The expectation values of the observables are given by the followings.
\begin{eqnarray}\label{expectationvalues}
 \langle \hat{x}_1\rangle &=& \langle \hat{x}_2\rangle =\langle \hat{p}_1\rangle =\langle \hat{p}_2\rangle =0. \label{x1p1x2p2expectation}\\
 \langle \hat{x}_1^2\rangle &=& \frac{\Lambda_{22,r}}{2\Delta_r},\; \langle \hat{x}_2^2\rangle = \frac{\Lambda_{11,r}}{2\Delta_r}, \; \langle \hat{x}_1 \hat{x}_2 \rangle = - \frac{\Lambda_{12,r}}{2\Delta_r}. \\
 \langle \hat{p}_1^2\rangle &=& \frac{1}{2\Delta_r} \left( \Lambda_{22,r} \vert \Lambda_{11} \vert^2 + \Lambda_{11,r}  \Lambda_{12}^2 -2i \Lambda_{12,i} \Lambda_{12,r} \Lambda_{11}^* \right). \\
 \langle \hat{p}_2^2\rangle &=& \frac{1}{2\Delta_r} \left( \Lambda_{11,r} \vert \Lambda_{22} \vert^2 + \Lambda_{22,r}  \Lambda_{12}^2 -2i \Lambda_{12,i} \Lambda_{12,r} \Lambda_{22}^* \right).\\
 \langle \hat{p}_1 \hat{p}_2 \rangle &=&   \Lambda_{12} - \Lambda_{11} \Lambda_{12} \langle \hat{x}_1^2 \rangle - \Lambda_{22} \Lambda_{12} \langle \hat{x}_2^2 \rangle - (\Lambda_{11} \Lambda_{22} + \Lambda_{12}^2) \langle \hat{x}_1 \hat{x}_2 \rangle .\\
 \langle \left\{ \hat{x}_1,  \hat{p}_1 \right\} \rangle &=&   2i \Lambda_{11} \langle \hat{x}_1^2 \rangle + 2\Lambda_{12} \langle \hat{x}_1 \hat{x}_2  \rangle -1 .\\
 \langle \left\{ \hat{x}_2,  \hat{p}_2 \right\} \rangle &=&    2i \Lambda_{22} \langle \hat{x}_2^2 \rangle + 2\Lambda_{12} \langle \hat{x}_2 \hat{x}_1  \rangle -1 .\\
 \langle \hat{x}_1  \hat{p}_2  \rangle &=& i   \Lambda_{12} \langle \hat{x}_1^2 \rangle + \Lambda_{22} \langle \hat{x}_1 \hat{x}_2  \rangle  , \;
 \langle \hat{x}_2  \hat{p}_1  \rangle = i   \Lambda_{12} \langle \hat{x}_2^2 \rangle + \Lambda_{11} \langle \hat{x}_2 \hat{x}_1  \rangle  .\\
 \mbox{Where}\; \Delta_r &=& \Lambda_{11,r}\Lambda_{22,r} - \Lambda_{12,r}^2 .
\end{eqnarray}

Here the suffix $r$ and $i$ denotes the real and imaginary part of the time-dependent parameters. For example $\Lambda_{11}=\Lambda_{11,r} + i\Lambda_{11,i}$.\\
Now from ~\eqref{x1p1x2p2expectation}, we find that $\mathcal{V}_{\alpha\beta}$ in ~\eqref{valphabeta} simplifies to $\mathcal{V}_{\alpha\beta} =\frac{1}{2}\langle \{\hat{X}_\alpha,\hat{X}_\beta\}\rangle$, i.e.,
\begin{eqnarray}\label{V1122exact}
 \hat{V}_{\alpha\alpha}=\left(\begin{array}{cc}
\langle \hat{x}_\alpha^2 \rangle & \langle \frac{1}{2}\{ \hat{x}_\alpha,\hat{p}_\alpha \} \rangle \\
\langle \frac{1}{2}\{ \hat{x}_\alpha,\hat{p}_\alpha \} \rangle \rangle  & \langle \hat{p}_\alpha^2 \rangle
\end{array}
\right), \;
\hat{V}_{12}= \left(\begin{array}{cc}
\langle \hat{x}_1\hat{x}_2 \rangle & \langle \hat{x}_1\hat{p}_2  \rangle \\
\langle \hat{p}_1\hat{x}_2  \rangle  & \langle \hat{p}_1\hat{p}_2  \rangle
\end{array}
\right),\; \alpha=1,2.
\end{eqnarray}
Using the explicit forms of expectation values in ~\eqref{V1122exact},  the Simon's separability criterion  ~\eqref{simonscriterion} corresponding to the ground state of our system reads
\begin{eqnarray}
 \vert \Lambda_{12} \vert^2 \Delta_r^2 - ( \Lambda_{12,r}^4 + 2\Lambda_{12,i}^2 \Lambda_{11,r} \Lambda_{22,r}) \Delta_r 
 + 2 \Lambda_{12,i}^2 (2 \Lambda_{12,r}^4 - \Lambda_{11,r} \Lambda_{22,r} \Lambda_{12,i}^2) \nonumber \\
 + 3\Lambda_{11,r}\Lambda_{22,r} (\Lambda_{11,r}^2 \Lambda_{22,r}^2 - 2 \Lambda_{12,r}^4) + 3 \Lambda_{11,r}^2 \Lambda_{22,r}^2 \Lambda_{12,r}^2 \ge 0.
\end{eqnarray}

\section{Conclusions}
The exact solutions of the time-dependent Schr\"{o}dinger equation (SE) for an anisotropic harmonic oscillator are constructed with the help of Lewis-Riesenfeld invariant method. We had considered an invariant operator, which is quadratic in co-ordinates and momentum. The coupled first order differential equations of the coefficients in the linear combination of the constituent operators in the invariant operator are expressed in a closed matrix form. These coupled matrix equations can be solved in principle for a given set of time-dependent parameters (mass, field strength, frequencies, NC parameter - whichever is relevant). The quadratic form is expressed in a simpler form by diagonalizing the system keeping the intrinsic symplectic structure intact. The annihilation and creation operators are introduced. The eigenstates and eigenvalues are demonstrated in detail. Both the geometrical and dynamical phases are constructed, and thus  Schr\"{o}dinder equation is completely solved for our system.  The ground state for the bipartite system is demonstrated along with its separability criterion. \\
The results discussed in this paper can be directly used for various types of systems. In particular, for an anisotropic oscillator in noncommutative space (for both the spatial and momentum noncommutativity), charged  oscillator in a anisotropic magnetic field, for Chern-Simon model (in long wavelength limit) and of the similar types.
\section{Author Declarations}
The authors have no conflicts to disclose.
\section{Availability of data}
Data sharing is not applicable to this article as no new data were created or analyzed in this study.
\section{Acknowledgement}
This research was supported in part by the International Centre for Theoretical Sciences (ICTS) for the online program - Non-Hermitian Physics (code: ICTS/nhp2021/3).


\begin{thebibliography}{99} 
\bibitem{tdho1} A. Mostafazadeh, \textquotedblleft Time-dependent diffeomorphisms as quantum canonical transformations and the time-dependent harmonic oscillator. \textquotedblright, J. Phys. A: Math. Gen. {\bf 31}, 6495 (1998).
\bibitem{tdho2} G. S. Agarwal and S. Arun Kumar, \textquotedblleft Exact quantum-statistical dynamics of an oscillator with time-dependent frequency and generation of nonclassical states. \textquotedblright, Phys. Rev. Lett. {\bf 67}, 3665 (1991).
\bibitem{tdho3} T. Beus, J. F. Van Huele and M. Berrondo,  \textquotedblleft Quantum manipulation through finite fluctuations for a generalized parametric oscillator using a lie algebra representation. \textquotedblright, Phys. Scr. {\bf 96}, 075103 (2021).
\bibitem{tdho4}F. Vega, \textquotedblleft Oscillators in a (2+1)-dimensional noncommutative space.  \textquotedblright, J. Math. Phys. {\bf 55} 032105 (2014).
\bibitem{tdho5}D. M. Tibaduiza, L. Pires and C. Farina, \textquotedblleft Time-dependent quantum harmonic oscillator: a continuous route from adiabatic to sudden changes. \textquotedblright, J. Phys. B: At. Mol. Opt. Phys. {\bf 54}, 205401 (2021).
\bibitem{tdho6} F. Soto-Eguibar, F. A. Asenjo, S. A. Hojman and H. M. Moya-Cessa, \textquotedblleft Bohm potential for the time dependent harmonic oscillator . \textquotedblright, J. Math. Phys. {\bf 62}, 122103 (2021).
\bibitem{ncs1}R. Banerjee, B. Chakraborty, S. Ghosh, P. Mukherjee and S. Samanta, \textquotedblleft Topics in Noncommutative Geometry Inspired Physics. \textquotedblright, Found Phys {\bf 39}, 1297 (2009). 
\bibitem{ncs2}S. Ghosh, S. Gangopadhyay
and P. K. Panigrahi, \textquotedblleft Noncommutative quantum cosmology with perfect fluid. \textquotedblright,  Modern Physics Letters A {\bf 37}, 2250009 (2022). 
\bibitem{ncs3}R. J. Szabo, \textquotedblleft Quantum field theory on noncommutative spaces. \textquotedblright , Physics Reports {\bf 378}, 207-299 (2003).
\bibitem{ncs4}M. R. Douglas and N. A. Nekrasov, \textquotedblleft Noncommutative field theory. \textquotedblright, Rev. Mod. Phys. {\bf 73} 977 (2001).
 \bibitem{qg1}A. Ashtekar and R. Geroch, \textquotedblleft Quantum theory of gravitation. \textquotedblright,  Rep. Prog. Phys. {\bf 37}, 1211 (1974).
  \bibitem{qg2}K.S. Stelle, \textquotedblleft The unification of quantum gravity. \textquotedblright, Nuclear Physics B - Proceedings Supplements {\bf 88}, 3-9 (2000).
  \bibitem{qg3}C. Pfeifer and J. J. Relancio, \textquotedblleft Deformed relativistic kinematics on curved spacetime: a geometric approach. \textquotedblright, Eur. Phys. J. C {\bf 82}, 150 (2022). 
\bibitem{qg4} S. Liberati, \textquotedblleft Tests of Lorentz invariance: a 2013 update. \textquotedblright,  Class. Quantum Gravity {\bf 30}, 133001 (2013). 
\bibitem{qg5}N. Seiberg and  E. Witten, \textquotedblleft String theory and noncommutative geometry. \textquotedblright, Journal of High Energy Physics {\bf 09}, (1999).

\bibitem{qg6}J. M. Romero, J. D. Vergara, and J. A. Santiago, \textquotedblleft Noncommutative spaces, the quantum of time, and Lorentz symmetry. \textquotedblright, Phys. Rev. D {\bf 75}, 065008 (2007).
\bibitem{qg7}C. L. Ching and W. K. Kg,  \textquotedblleft Deformed Gazeau-Klauder Schr\"{o}dinger cat states with modified commutation relations. \textquotedblright ,
Phys. Rev. D {\bf 100}, 085018 (2019).
\bibitem{qg8}J. F. G. Santos, \textquotedblleft Noncommutative phase-space effects in thermal diffusion of Gaussian states. \textquotedblright, J. Phys. A: Math. Theor. {\bf 52}, 405306 (2019).
\bibitem{qg9}P. Chattopadhyay, A. Mitra and G. Paul, \textquotedblleft Uncertainty Relations in Non-Commutative Space. \textquotedblright,
 Int. J. Theor. Phys. {\bf 58}, 2619-2631 (2019).
\bibitem{mcs1} L. Inzunza and M. S. Plyushchay, \textquotedblleft Conformal generation of an exotic rotationally invariant harmonic oscillator. \textquotedblright, Phys. Rev. D {\bf 103}, 106004 (2021).
\bibitem{mcs2}K. Biswas, J. P. Saha and P. Patra \textquotedblleft Squeezed coherent state for free falling Maxwell-Chern-Simons model in long-wavelength limit. \textquotedblright, Indian J Phys {\bf 95}, 647-655 (2021).
\bibitem{mcs3} P. A. Horv\'{a}thy and M. S. Plyushchay,\textquotedblleft Anyon wave equations and the noncommutative plane. \textquotedblright, Physics Letters B {\bf 595} 547-555 (2004).
\bibitem{mcs4}J. Han, H. Huh and J. Seok \textquotedblleft Chern-Simons limit of the standing wave solutions for the Schr\"{o}dinger equations coupled with a neutral scalar field. \textquotedblright, Journal of Functional Analysis {\bf 266}, 318-342 (2014).
\bibitem{mcs5} A. Lerda, \textquotedblleft  Anyons in a Magnetic Field.   \textquotedblright, Anyons. Lecture Notes in Physics Monographs {\bf 14}, Springer, Berlin, Heidelberg (1992). 
\bibitem{mcs6} H. Bartolomei, et al. \textquotedblleft Fractional statistics in anyon collisions. \textquotedblright, Science {\bf 368}, 173-177 (2020).
\bibitem{mcs7} T. Girardot, \textquotedblleft Average field approximation for almost bosonic anyons in a magnetic field. \textquotedblright,  J. Math. Phys. {\bf 61}, 071901 (2020).
\bibitem{mcs8}R. J. Szabo and M. Tierz, \textquotedblleft Chern-Simons matrix models, two-dimensional Yang-Mills theory and the Sutherland model \textquotedblright, J. Phys. A: Math. Theor. {\bf 43} 265401 (2010).
\bibitem{tdem1}D. Wing et al., \textquotedblleft Comparing time-dependent density functional theory with many-body perturbation theory for semiconductors: Screened range-separated hybrids and the GW plus Bethe-Salpeter approach.\textquotedblright, Phys. Rev. Materials {\bf 3}, 064603 (2019).
\bibitem{tdem2}M. G. Silveirinha, N. Engheta, \textquotedblleft Transformation electronics: Tailoring the effective mass of
electrons.\textquotedblright, Phys. Rev. B {\bf 86} 161104(R) (2012).
\bibitem{tdem3} C. Quesne and V. M. Tkachuk, \textquotedblleft Deformed algebras, position-dependent effective masses and
curved spaces: an exactly solvable Coulomb problem \textquotedblright, J. Phys. A: Math. Gen. {\bf 37} 4267
(2004).
\bibitem{tdem5} M. S. Cunha, C. R. Muniz, H. R. Christiansen and V. B. Bezerra, \textquotedblleft Relativistic Landau levels in
the rotating cosmic string spacetime \textquotedblright, Eur. Phys. J. C {\bf 76} 512 (2016).
\bibitem{tdem6}S. Zapperi, C. Castellano, F. Colaiori and G. Durin, \textquotedblleft Signature of effective mass in crackling-noise
asymmetry \textquotedblright, Nat. Phys. {\bf 1} 46 (2005).
\bibitem{tdem7}A. S. Halberg,\textquotedblleft  Quantum systems with effective and time-dependent masses: form preserving
transformations and reality conditions \textquotedblright, Open Phys. {\bf 3} 591 (2005).
\bibitem{tdem8} K. Biswas, J. P. Saha and P. Patra, \textquotedblleft On the position-dependent effective mass Hamiltonian. \textquotedblright,
Eur. Phys. J. Plus {\bf 135} 457 (2020).
\bibitem{cs1}M. Pitkin, S. Reid, S. Rowan and J. Hough, \textquotedblleft Gravitational Wave Detection by Interferometry (Ground and Space)\textquotedblright, Living Rev. Relativ. {\bf 14} 5 (2011). 
\bibitem{cs2} C. L. Ching and W. K. Ng, \textquotedblleft Deformed Gazeau-Klauder Schr\"{o}dinger cat states with modified commutation relations\textquotedblright,
Phys. Rev. D {\bf 100} 085018 (2019).
\bibitem{cs3}J. F. G. Santos, \textquotedblleft Noncommutative phase-space effects in thermal diffusion of Gaussian states\textquotedblright, J. Phys. A: Math. Theor. {\bf 52} 405306 (2019).
\bibitem{cs4}P. Chattopadhyay, A. Mitra and G. Paul, \textquotedblleft Uncertainty Relations in Non-Commutative Space\textquotedblright,
 Int. J. Theor. Phys. \textbf{58} 2619-2631 (2019).
\bibitem{cs5}U. D. Jentschura, \textquotedblleft Gravitational effects in {\it g}-factor measurements and high-precision spectroscopy: Limits of Einstein's equivalence principle\textquotedblright, Phys. Rev. A {\bf 98} 032508 (2018).
\bibitem{cs6}D. Buono, G. Nocerino, V. D'Auria, A. Porzio, S. Olivares, and M. G. A. Paris, \textquotedblleft Quantum characterization of bipartite Gaussian states \textquotedblright, Journal of the Optical Society of America B {\bf 27} A110-A118 (2010).
\bibitem{cs7}S. Ma, M. J. Woolley, X. Jia and J. Zhang, \textquotedblleft Preparation of bipartite bound entangled Gaussian states in quantum optics\textquotedblleft, Phys. Rev. A {\bf 100}, 022309 (2019).

\bibitem{cs8}S. M. Vermeulen et al, \textquotedblleft An experiment for observing quantum gravity phenomena using twin table-top 3D interferometers \textquotedblright  Class. Quantum Grav. {\bf 38} 085008 (2021).
\bibitem{cs9}J. A. de F. Pacheco, S. Carneiro and J. C. Fabris, \textquotedblleft Gravitational waves from binary axionic black holes\textquotedblright, Eur. Phys. J. C {\bf 79} 426 (2019). 

\bibitem{cs10}M. J. Koop and L. S. Finn, \textquotedblleft Physical response of light-time gravitational wave detectors\textquotedblright, Phys. Rev. D {\bf 90} 062002 (2014).
\bibitem{csncs1}A. Muhuri, D. Sinha and S. Ghosh, \textquotedblleft Entanglement induced by noncommutativity: anisotropic harmonic oscillator in noncommutative space\textquotedblright, Eur. Phys. J. Plus {\bf 136} 35 (2021). 
\bibitem{csncs2}M. C. Eser and M. Riza, \textquotedblleft Energy corrections due to the noncommutative phase-space of the charged isotropic harmonic oscillator in a uniform magnetic field in 3D\textquotedblright, Phys. Scr. {\bf 96} 085201 (2021).
\bibitem{csncs3}B. Lin, J. Xu
and T. Heng, \textquotedblleft Induced entanglement entropy of harmonic oscillators in non-commutative phase space \textquotedblright,  Modern Physics Letters A {\bf 34}  1950268 (2019).
\bibitem{csncs4}C. Bastos, A. E. Bernardini, O. Bertolami, N. C. Dias and J. N. Prata, \textquotedblleft Entanglement due to noncommutativity in phase space\textquotedblright, Phys. Rev. D {\bf 88} 085013 (2013).

\bibitem{Lewis1}H. R. Lewis, Jr., \textquotedblleft Classical and Quantum Systems with Time-Dependent Harmonic-Oscillator-Type Hamiltonians. \textquotedblright, Phys. Rev. Lett. {\bf 18}, 636  (1967).
\bibitem{Lewis2}H. R. Lewis Jr., \textquotedblleft Class of Exact Invariants for Classical and Quantum Time-Dependent Harmonic Oscillators. \textquotedblright, J. Math. Phys. {\bf 9}, 1976 (1968).
\bibitem{Lewis3}H. R. Lewis Jr. and W. B. Riesenfeld, \textquotedblleft An Exact Quantum Theory of the Time-Dependent Harmonic Oscillator and of a Charged Particle in a Time-Dependent Electromagnetic Field. \textquotedblright, J. Math. Phys. {\bf 10}, 1458 (1969).
\bibitem{Lewis4} M. A. de Ponte, P. M. C\^{o}nsoli, and M. H. Y. Moussa, \textquotedblleft Method for the construction of the Lewis-Riesenfeld time-dependent invariants and their eigenvalue equations.\textquotedblright,
Phys. Rev. A {\bf 98}, 032102 (2018).

\bibitem{Lewis5}  L. Lawson, L. Gouba and G. Y. Avossevou, \textquotedblleft Two-dimensional noncommutative gravitational quantum well. \textquotedblright, J. Phys. A: Math. Theor. {\bf 50} 475202 (2017).
\bibitem{Lewis6}P. Patra, J. P. Saha and K. Biswas, \textquotedblleft Squeezed coherent states for gravitational well in noncommutative space. \textquotedblright, Indian J. Phys, {\bf 96}, 309-315 (2022).
\bibitem{Lewis7}L. M. Lawson,  G. Y. H. Avossevou and  L. Gouba, \textquotedblleft Lewis-Riesenfeld quantization and SU(1,1) coherent states for 2D damped harmonic oscillator. \textquotedblright, J. Math. Phys. {\bf 59}, 112101 (2018).
\bibitem{anisotropic1}L. Qiong-Gui, \textquotedblleft Anisotropic Harmonic Oscillator in a Static Electromagnetic Field\textquotedblright, Commun. Theor. Phys. {\bf 38}, 667 (2002).
\bibitem{sp1} Arvind, N. Mukunda, R. Simon \textquotedblleft The real symplectic groups in quantum mechanics and optics \textquotedblright, Pramana - J Phys  {\bf 45 } 471-497 (1995).
\bibitem{phase1}M. V. Berry, \textquotedblleft Quantal phase factors accompanying adiabatic changes. \textquotedblright, Proc. R. Soc. London A {\bf 392}, 45 (1984).
\bibitem{phase2} B. Simon, \textquotedblleft Holonomy, the Quantum Adiabatic Theorem, and Berry's Phase\textquotedblright, Phys. Rev. Lett. {\bf 51}, 2167 (1983).
\bibitem{phase3} Y. Aharonov and J. Anandan,\textquotedblleft Phase change during a cyclic quantum evolution\textquotedblright, Phys. Rev. Lett. {\bf 58}, 1593 (1987).
\bibitem{phase4}G. Fiore and L. Gouba, \textquotedblleft Class of invariants for the two-dimensional time-dependent
Landau problem and harmonic oscillator in a magnetic field. \textquotedblright, J. Math. Phys {\bf 52}, 103509 (2011).
\bibitem{sep1}R. Simon, \textquotedblleft Peres-Horodecki Separability Criterion for Continuous Variable Systems. \textquotedblright, Phys. Rev. Lett.  {\bf 84}, 2726 (2000).
\bibitem{sep2}A. Peres, \textquotedblleft Separability Criterion for Density Matrices.\textquotedblright, Phys. Rev. Lett.   {\bf 77},  1413 (1996).
\bibitem{sep3}P. Horodecki, \textquotedblleft Separability criterion and inseparable mixed states with positive partial transposition.\textquotedblright, Physics Letters A {\bf 232}, 333-339 (1997).
\bibitem{sep4}J. Eisert, T. Tyc, T. Rudolph and B. C. Sanders, \textquotedblleft Gaussian Quantum Marginal Problem. \textquotedblright,  Commun. Math. Phys. {\bf 280}, 263-280 (2008).
\bibitem{sep5}V. V. Dodonov, \textquotedblleft Invariant Quantum States of Quadratic Hamiltonians. \textquotedblright, Entropy {\bf 2031}, 634 (2021).
\bibitem{sep6}M. Moshinsky and P. Winternitz, \textquotedblleft Quadratic Hamiltonians in phase space and their eigenstates \textquotedblright, J. Math. Phys. {\bf 21} 1667 (1980).
\end{thebibliography}
 \end{document}